\documentclass[final,3p,times]{elsarticle}
\usepackage{amssymb}
\usepackage{amssymb}
\usepackage[]{amsmath}
\usepackage{graphics}
\usepackage{amssymb}
\usepackage[]{amsmath}
\usepackage{amsthm}
\usepackage{setspace}
\usepackage{epsfig}
\usepackage{subfigure}
\usepackage{booktabs}

\biboptions{numbers,sort&compress}
\usepackage[dvipdfm,colorlinks,linkcolor=red,anchorcolor=blue,citecolor=green]{hyperref}

\usepackage{amsmath}
\usepackage{times}
\usepackage{anysize}
\marginsize{2.5cm}{2.5cm}{1cm}{2cm}
\selectfont

 \journal {}

\begin{document}

\begin{frontmatter}

%% Title, authors and addresses

%% use the tnoteref command within \title for footnotes;
%% use the tnotetext command for the associated footnote;
%% use the fnref command within \author or \address for footnotes;
%% use the fntext command for the associated footnote;
%% use the corref command within \author for corresponding author footnotes;
%% use the cortext command for the associated footnote;
%% use the ead command for the email address,
%% and the form \ead[url] for the home page:
%%
%% \title{Title\tnoteref{label1}}
%% \tnotetext[label1]{}
%% \author{Name\corref{cor1}\fnref{label2}}
%% \ead{email address}
%% \ead[url]{home page}
%% \fntext[label2]{}
%% \cortext[cor1]{}
%% \address{Address\fnref{label3}}
%% \fntext[label3]{}

\title{Nonlocal symmetries and conservation laws of the coupled Hirota equation}

\author{ Xiangpeng Xin \fnref{label2} \corref{cor1}}
%\author{ Yong Chen \fnref{label2} }
\ead{xinxiangpeng@lcu.edu.cn}

\cortext[cor1]{Corresponding author. School of Mathematical Sciences, Liaocheng University, Liaocheng 252059, People's Republic of China}

\address[label2]{ School of Mathematical Sciences, Liaocheng University, Liaocheng 252059, People's Republic of China}

\begin{abstract}
%% Text of abstract
Using the lax pair, nonlocal symmetry of the coupled Hirota equation is obtained. By introducing an appropriate auxiliary dependent variable the nonlocal symmetry is successfully localized to a Lie point symmetry. For the closed prolongation, one-dimensional optimal systems and nonlocal conservation laws of the coupled Hirota equation are studied.

\end{abstract}

\end{frontmatter}

%%
%% Start line numbering here if you want
%%
% \linenumbers

%% main text
\section{Introduction}
%In nonlinear science, especially in integrable systems, one of the most important and essential task to
%find symmetries, conservation laws, symmetry groups, symmetry reductions and group invariant solutions of partial differential equations.

The Lie symmetries\cite{Lie1,Ovsiannikov1,Ibragimov1,Bluman1,Chen1,Liu1} and their various generalizations have become an important subject in mathematics and physics. One can reduce the dimensions of partial differential equations(PDEs) and proceed to construct analytical solutions by using classical or non-classical Lie symmetries. However, with all its importance and power, the traditional Lie approach does not provide all the answers to mounting challenges of the modern nonlinear physics. In the 80s of the last century, there exist so-called nonlocal symmetries which entered the literature largely through the work of Olver\cite{Olver1}. Compared with the local symmetries, little importance is attached to the existence and applications of the nonlocal ones. The reason lies in that nonlocal symmetries are difficult to find and similarity reductions cannot be directly calculated. Many researchers\cite{Akhatov,Bluman,Lou} have done a lot of work in this area, references\cite{Galas1,Lou1,Hu1} give a direct way to solve this problem which so-called localization method of nonlocal symmetries. I.e. the original system is prolonged to a larger system such that the nonlocal symmetry of the original model becomes a local one of the prolonged system. When we get the Lie symmetries of prolonged system, there will correspond a family of group-invariant solutions. Since there are almost always an infinite number of such subgroups, it is not usually feasible to list all possible group-invariant solutions to the system. Hence, the optimal system\cite{Ovsiannikov1,David1,Qu1,Liu2,Hu2} of group-invariant solutions should be constructed.

Conservation laws are used for the development of appropriate numerical methods and for mathematical analysis, in particular, existence,uniqueness and stability analysis. It can lead to some new integrable systems via reciprocal transformation. The famous Noether's theorem\cite{Noether1} provides a systematic way of determining conservation laws, for Euler-Lagrange differential equations, to each Noether symmetry associated with the Lagrangian there corresponds a conservation law which can be determined explicitly by a formula. But this theorem relies on the availability of classical Lagrangians. To find conservation laws of differential equations without classical Lagrangians, researchers have made various generalizations of Noether's theorem. Steudel\cite{Steudel1} writes a conservation law in characteristic form, where the characteristics are the multipliers of the differential equations. In order to determine a conservation law one has to also find the related characteristics. Anco and Bluman\cite{Anco1} provides formulae for finding conservation laws for known characteristics. Infinitely many nonlocal conservation laws for(1+1)-dimensional evolution equations are revealed by Lou\cite{Lou2}. Symmetry considerations for PDEs were incorporated by Ibragimov\cite{Ibragimov2} which can be computed by a formula.

This paper is arranged as follows: In Sec.\ref{nonlocal}, the nonlocal symmetries of the coupled Hirota equation are obtained by using the Lax pair. In Sec.\ref{liepoint}, we transform the nonlocal symmetries into Lie point symmetries. Then, the finite symmetry transformations are obtained by solving the initial value problem. In Sec.\ref{optimal}, an optimal system is constructed to classify the group-invariant solutions of the coupled Hirota equation. In Sec.\ref{reduction}, based on the symmetries of prolonged system, nonlocal conservation laws of the coupled Hirota equation are given out. Finally, some conclusions and discussions are given in Sec.\ref{conclusion}.

\section{Nonlocal symmetries of the coupled Hirota equation}\label{nonlocal}

The well-known the Hirota equation\cite{Ablowitz1,Ankiewicz1} reads
\begin{equation}\label{ch-0}
iu_t  + \alpha (u_{xx}  + 2\left| u \right|^2 u) + i\beta (u_{xxx}  + 6\left| u \right|^2 u_x ) = 0,
\end{equation}
$\alpha,\beta$ are real constants. Eq.(\ref{ch-0}) is the third flow of the nonlinear Schr\"{o}dinger (NLS)hierarchy which can be used to describe many kinds of nonlinear phenomenas or mechanisms in the fields of physics,optical fibers,electric communication and other engineering sciences. Eq.(\ref{ch-0})reduces to NLS equation when $\alpha=1,\beta=0$.

In this section, we shall consider the coupled Hirota equations
\begin{equation}\label{ch-1}
\begin{array}{l}
 iu_t  + \alpha (u_{xx}  - 2u^2 v) + i\beta (u_{xxx}  - 6uvu_x ) = 0, \\
 iv_t  - \alpha (v_{xx}  - 2v^2 u) + i\beta (v_{xxx}  - 6uvv_x ) = 0, \\
 \end{array}
\end{equation}
Eqs.(\ref{ch-1})are reduced to the Eq.(\ref{ch-0})when $u=-v^*$, and $*$ denotes the complex conjugate.

The Lax pair of Eq.(\ref{ch-1}) has been obtained in\cite{Zhang1}
\begin{equation}\label{ch-2}
\Phi _x  = U\Phi ,U = \left( {\begin{array}{*{20}c}
   { - i\lambda } & u  \\
   v & {i\lambda }  \\
\end{array}} \right)
\end{equation}
and
\begin{equation}\label{ch-3}
\Phi _t  = V\Phi ,V = \left( {\begin{array}{*{20}c}
   a & b  \\
   c & { - a}  \\
\end{array}} \right)
\end{equation}
with
\begin{equation}\label{ch-3-1}
\begin{array}{l}
 a =  - 4\beta i\lambda ^3  - 2\alpha i\lambda ^2  - 2\beta iuv\lambda  - \alpha iuv + \beta (vu_x  - uv_x ), \\
 b = 4\beta u\lambda ^2  + (2\beta iu_x  + 2\alpha u)\lambda  + \alpha iu_x  - \beta (u_{xx}  - 2u^2 v), \\
 c = 4\beta v\lambda ^2  - (2\beta iv_x  - 2\alpha v)\lambda  - \alpha iv_x  - \beta (v_{xx}  - 2v^2 u). \\
 \end{array}
\end{equation}
where $u$ and $v$ are two potentials, the spectral parameter $\lambda$ is an arbitrary complex constant and its eigenfunction is  $\Phi =(\phi ,\psi)^T $.

To seek for the nonlocal symmetries, we adopt a direct method\cite{Xin1}. First of all, the symmetries $\sigma_1,\sigma_2$ of the coupled Hirota equations are defined as solutions of their linearized equations
\begin{equation}\label{ch-4}
\begin{array}{l}
 \sigma _{1t}  - \alpha \sigma _{1xx} i + 4i\alpha \sigma _1 uv + 2i\alpha u^2 \sigma _2  + \beta \sigma _{1xxx}  - 6\beta \sigma _1 vu_x  - 6\beta u\sigma _2 u_x  - 6\beta uv\sigma _{1x}  = 0, \\
 \sigma _{2t}  + \alpha \sigma _{2xx} i - 4i\alpha uv\sigma _2  - 2i\alpha v^2 \sigma _1  + \beta \sigma _{2xxx}  - 6\beta \sigma _2 uv_x  - 6\beta uv\sigma _{2x}  - 6\beta v\sigma _1 v_x  = 0, \\
 \end{array}
\end{equation}
which means equation (\ref{ch-1}) is form invariant under the infinitesimal transformations
\begin{equation}\label{ch-5}
\begin{array}{l}
 u \to u + \epsilon \sigma _1,  \\
 v \to v + \epsilon \sigma _2,  \\
 \end{array}
\end{equation}
with the infinitesimal parameter $\epsilon $.

The symmetry can be written as
\begin{equation}\label{ch-6}
\begin{array}{l}
 \sigma _1 =X(x,t,u,v,\phi ,\psi )u_x  + T(x,t,u,v,\phi ,\psi )u_t  - U(x,t,u,v,\phi ,\psi ), \\
 \sigma _2 =X(x,t,u,v,\phi ,\psi )v_x  + T(x,t,u,v,\phi ,\psi )v_t  - V(x,t,u,v,\phi ,\psi ), \\
 \end{array}
\end{equation}

Here, $X,T,U,V$ dependent on the auxiliary variables $\phi$ and $\psi$, so one may obtain some different results from Lie symmetries. Substituting Eq.(\ref{ch-6}) into Eq.(\ref{ch-4}) and eliminating $u_t,v_t,\phi_x,\phi_t,\psi_x,\psi_t $ in terms of the closed system, it yields a system of determining equations for the functions $X,T,U,V$, which can be solved by virtue of Maple to give
\begin{equation}\label{ch-7}
\begin{array}{l}
 X = \frac{{c_1 x}}{3} + \frac{{2\alpha ^2 c_1 t}}{{9\beta }} + c_3 , \\
 T(x,t,u,v,\phi ,\psi ) = c_1 t + c_2 , \\
 U(x,t,u,v,\phi ,\psi ) = \frac{{i\alpha c_1 ux}}{{9\beta }} + c_5 u + c_4 \phi ^2 , \\
 V(x,t,u,v,\phi ,\psi ) = \frac{{(( - 6c_1 {- 9c}_5 )v + 9c_4 \psi ^2 )\beta  - i\alpha vc_1 x}}{{9\beta }} \\
 \end{array}
\end{equation}
where $c_i (i = 1, . . . ,5)$ are five arbitrary constants and $i^2=-1$. It can be seen from the results(\ref{ch-7}), the results contain a nonlocal symmetry when $c_4\neq 0$.

\section{Localization of the nonlocal symmetry}\label{liepoint}

As we all know, the nonlocal symmetries cannot be used to construct explicit solutions directly. Hence, one need to transform the nonlocal symmetries into local ones\cite{Galas1,Lou1,Hu1}. In this section, a related system which possesses a Lie point symmetry that is equivalent to the nonlocal symmetry will be found.

For simplicity, we let $c_1=c_2=c_3=c_5=0,c_4=-1$ in formula (\ref{ch-7}), i.e.,
\begin{equation}\label{ch-8}
\begin{array}{l}
 \sigma _1  = \phi ^2 , \\
 \sigma _2  = \psi ^2 . \\
 \end{array}
\end{equation}

To localize the nonlocal symmetry (\ref{ch-8}), we have to solve the following linearized equations
\begin{equation}\label{ch-9}
\begin{array}{l}
 \sigma _{3x}  + i\lambda \sigma _3  - \sigma _1 \phi  - u\sigma _4  = 0, \\
 \sigma _{4x}  - \sigma _2 \varphi  - v\sigma _3  - i\lambda \sigma _4  = 0, \\
 \end{array}
\end{equation}
which means that Eqs.(\ref{ch-2}) invariant under the infinitesimal transformations
\begin{equation}\label{ch-9-1}
\begin{array}{l}
 \phi \to \phi + \epsilon \sigma _3,  \\
 \psi \to \psi + \epsilon \sigma _4,  \\
 \end{array}
\end{equation}
with $\sigma_1,\sigma_2$ given by (\ref{ch-8}). It is not difficult to verify that the solutions of (\ref{ch-9}) have the following forms
\begin{equation}\label{ch-10}
\sigma _3  = \phi f,~~~~\sigma _4  =\psi f,
\end{equation}
where $f$ is given by
\begin{equation}\label{ch-11}
\begin{array}{l}
 f_x  = \phi \psi ,\\
 f_t  =  - \beta \phi ^2 v_x  + 12\beta \lambda ^2 \phi \psi  + 4\lambda \alpha \phi \psi  - \beta \psi ^2 u_x  + 2\beta \phi \psi uv +4i\lambda \beta \psi ^2 u - 4i\lambda \beta \phi ^2 v + i\alpha \psi ^2 u - i\alpha \phi ^2 v. \\
 \end{array}
\end{equation}

It is easy to obtain the following result
\begin{equation}\label{ch-12}
\sigma_5=\sigma_f=f^2.
\end{equation}

The results (\ref{ch-10}) and (\ref{ch-12}) show that the nonlocal symmetry (\ref{ch-8}) in the original space $\{x,t,u,v\}$ has been successfully localized to a Lie point symmetry in the enlarged space $\{x,t,u,v,\phi,\psi,f\}$ with the vector form
\begin{equation}\label{ch-13}
V_1  = \phi ^2 \frac{\partial }{{\partial u}} + \psi ^2 \frac{\partial }{{\partial v}} + \phi f\frac{\partial }{{\partial \phi }} + \psi f\frac{\partial }{{\partial \psi }} + f^2 \frac{\partial }{{\partial f}}.
\end{equation}

After succeeding in making the nonlocal symmetry(\ref{ch-8}) equivalent to Lie point symmetry (\ref{ch-13}) of the related prolonged system, we can construct the explicit solutions naturally by Lie group theory. With the Lie point symmetry(\ref{ch-13}), by solving the following initial value problem
\begin{equation}\label{ch-14}
\begin{array}{l}
 \frac{{d\overline u }}{{d\epsilon }} = \phi ^2 ,~~~~\overline u |_{\epsilon  = 0}  = u, \\
 \frac{{d\overline v }}{{d\epsilon }} = \psi ^2 ,~~~~\overline v |_{\epsilon  = 0}  = v, \\
 \frac{{d\overline \phi  }}{{d\epsilon }} = \phi f,~~~~\overline \phi  |_{\epsilon  = 0}  = \phi , \\
 \frac{{d\overline \psi  }}{{d\epsilon }} = \psi f,~~~~\overline \psi  |_{\epsilon  = 0}  = \psi , \\
 \frac{{d\overline f }}{{d\epsilon }} = f^2 ,~~~~\overline f |_{\epsilon  = 0}  = f, \\
 \end{array}
\end{equation}
the finite symmetry transformation can be calculated as
\begin{equation}\label{ch-15}
\overline u  = \frac{{\epsilon fu - \varepsilon \phi ^2  - u}}{{\epsilon f - 1}},\overline v  = \frac{{\epsilon fv - \epsilon \psi ^2  - v}}{{\epsilon f - 1}},\overline \phi   = \frac{\phi }{{\varepsilon f - 1}},\overline \psi   = \frac{\psi }{{\epsilon f - 1}},\overline f  = \frac{f}{{\epsilon f - 1}},
\end{equation}

For a given solution $u,v,\phi,\psi,f$ of Eqs.(\ref{ch-15}), above finite symmetry transformation will arrive at another solution $\bar u,\bar v$. To search for more similarity reductions of Eqs.(\ref{ch-1}), one should consider Lie point symmetries of the whole prolonged system and assume the vector of the symmetries has the form
\begin{equation}\label{ch-16}
 V = X\frac{\partial }{{\partial x}} + T\frac{\partial }{{\partial t}} + U\frac{\partial }{{\partial u}}+ V\frac{\partial }{{\partial v}} + P \frac{\partial }{{\partial p }}  + Q \frac{\partial }{{\partial q }}+ F \frac{\partial }{{\partial f }} ,
\end{equation}
which means that the closed system is invariant under the infinitesimal transformations
\begin{center}
$(x,t,u,p,q,f )  \to (x + \epsilon X,t + \epsilon T,u + \epsilon U, \phi + \epsilon P,\psi + \epsilon Q,f + \epsilon F)$,
\end{center}
with
\begin{equation}\label{ch-18}
\begin{array}{l}
 \sigma _1  = X(x,t,u,v,\phi,\psi,f)u_x  + T(x,t,u,v,\phi,\psi,f)u_t  - U(x,t,u,v,\phi,\psi,f), \\
 \sigma _2  = X(x,t,u,v,\phi,\psi,f)v_x  + T(x,t,u,v,\phi,\psi,f)v_t  - V(x,t,u,v,\phi,\psi,f), \\
 \sigma _3  = X(x,t,u,v,\phi,\psi,f)\phi_x  + T(x,t,u,v,\phi,\psi,f)\phi_t  - P(x,t,u,v,\phi,\psi,f), \\
 \sigma _4  = X(x,t,u,v,\phi,\psi,f)\psi_x  + T(x,t,u,v,\phi,\psi,f)\psi_t  - Q(x,t,u,v,\phi,\psi,f), \\
 \sigma _5  = X(x,t,u,v,\phi,\psi,f)f_x  + T(x,t,u,v,\phi,\psi,f)f_t  - F(x,t,u,v,\phi,\psi,f). \\
 \end{array}
\end{equation}
with $\sigma _1,\sigma _2,\sigma _3,\sigma _4,\sigma _5$ satisfy the linearized equations of Eqs.(\ref{ch-1},\ref{ch-2},\ref{ch-3},\ref{ch-11}). Omitted here because of the formulas are long-winded.

Substituting Eq.(\ref{ch-18}) into linearized equations and eliminating $u_t ,v_t,\phi_x,\phi_t,\psi_x,\psi_t,f_x,f_t$ in terms of the closed system, one arrive at a system of determining equations for the functions $X,T,U,V,P,Q,$ and $F$, which can be solved by using Maple to give
\begin{equation}\label{ch-19}
\begin{array}{l}
 X(x,t,u,v,\phi ,\psi ,f) = c_4 ,\\
 T(x,t,u,v,\phi ,\psi ,f) = c_3 , \\
 U(x,t,u,v,\phi ,\psi ,f) = c_2 \phi ^2  + c_1 u, \\
 V(x,t,u,v,\phi ,\psi ,f) = c_2 \psi ^2  - c_1 v, \\
 P (x,t,u,v,\phi ,\psi ,f) = \frac{{(2c_2 f + c_1  + c_5 )\phi }}{2}, \\
 Q (x,t,u,v,\phi ,\psi ,f) = \frac{{( - 2c_2 f + c_1  - c_5 )\psi }}{2}, \\
 F(x,t,u,v,\phi ,\psi ,f) = c_2 f^2  + c_5 f + c_6 , \\
 \end{array}
\end{equation}
where $c_i,~~i=1,2,...,6$ are arbitrary constants.

\section{Optimal system of the prolonged system}\label{optimal}

In general, to each $s$-parameter subgroup of the full symmetry group, there will correspond a family of group-invariant solutions. Because there are always an infinite number of subgroups, there is no need to list possible group-invariant solutions to the system. In this section, an optimal system of one-dimensional subalgebras of Eq.(\ref{ch-1}) by using the method presented in Refs.\cite{Ovsiannikov1,Olver1} will be constructed.

As it is said in Ref.\cite{Ovsiannikov1}, the problem of finding an optimal system of subgroups is equivalent to finding an optimal system of subalgebras. From Eqs.(\ref{ch-19}), the associated vector fields for the one-parameter Lie group of infinitesimal transformations are six generators given by
\begin{equation}\label{ch-20}
\begin{array}{l}
 v_1  = \frac{1}{2}\phi \frac{\partial }{{\partial \phi }} + \frac{1}{2}\psi \frac{\partial }{{\partial \psi }} + f\frac{\partial }{{\partial f}},v_2  = \varphi ^2 \frac{\partial }{{\partial u}} + \psi ^2 \frac{\partial }{{\partial v}} + \phi f\frac{\partial }{{\partial \psi }} + f^2 \frac{\partial }{{\partial f}}, \\
 v_3  = \frac{\partial }{{\partial f}},v_4  = u\frac{\partial }{{\partial u}} - v\frac{\partial }{{\partial v}} + \frac{1}{2}\phi \frac{\partial }{{\partial \phi }} - \frac{1}{2}\psi \frac{\partial }{{\partial \psi }},v_5  = \frac{\partial }{{\partial t}},v_6  = \frac{\partial }{{\partial x}}, \\
 \end{array}
\end{equation}

One can know that $v_4,v_5,v_6$ are the centers of the group through calculating, so we don't have to consider them. Following Ref.\cite{Ovsiannikov1}, two subalgebras $v_2$ and $v_1$ of a given Lie algebra are equivalent if one can find an element $g$ in the Lie group so that $Adg({v_1}) = {v_2}$ where $Adg$ is the adjoint representation of $g$ on $v$. Given a nonzero vector, for example,
\begin{center}
$V=a_1v_1+a_2v_2+a_3v_3$,
\end{center}
where $a_j,~~j=1,2,3$ are arbitrary constants. The key task is to simplify as many of the coefficients $a_i$ as possible though judicious applications of adjoint maps to $v$. In this way, one can get the following results in Table 1 where $\alpha$ is an arbitrary constant.

\begin{table}[htbp]
\begin{center}
\caption{Optimal Systems}
\begin{tabular}{lclclcl}
 \toprule
  Cases &   & Optimal systems  \\
 \midrule
(a1) & $a_1  \ne 0$, & $v_1$  \\
(a2) & $a_3  \ne 0$, & $v_3 $  \\
(a3) & $a_2 \ne 0 ,a_3 \ne 0,$ ,& $v_2+\alpha v_3$  \\
 \bottomrule
 \end{tabular}
\end{center}
\end{table}

\section{Nonlocal conservation law of C-H equations}\label{reduction}

In this section, we briefly present the notations and theorems used in this paper firstly. Consider a system $F\{x;u\}$ of $N$ partial differential equations of order $s$ with $n$ independent variables $x = (x^1,..., x^n)$ and m dependent variables
$u(x) = (u^1(x),..., u^m(x))$, given by
\begin{equation}\label{ch-20}
{F_\alpha}[u]=F_\alpha(x,u,u_{(1)},...,u_{(s)}) = 0,\alpha  = 1,...,N
\end{equation}
where $u_{(1)}$,...$u_{(s)}$ denote the collection of
all first, . . ., $s$th-order partial derivatives. $u_i=D_i(u),D_{(ij)}=D_jD_i(u),...$. Here
$D_i=\frac{\partial} {\partial x_i }+u_i \frac{\partial }{\partial u }+u_{ij} \frac{\partial} {\partial u_j }+...,~~i=1,2,...,n$.

\textbf{Definition 1}: A conservation law of PDE system (\ref{ch-20}) is a divergence expression
\begin{equation}\label{ch-21}
{D_i}{\Phi ^i}[u] = {D_1}{\Phi ^1}[u] + ... + {D_n}{\Phi ^n}[u] = 0
\end{equation}
holding for all solutions of PDE system (\ref{ch-20}).

It is easy to see that Eqs.(\ref{ch-11}) determine a nonlocal conservation law, i.e.
\begin{equation}\label{ch-21-1}
D_t (f_x ) + D_x ( - f_t ) = 0.
\end{equation}

\textbf{Definition 2}:\cite{Ibragimov2}. The adjoint equations of Eq. (1) is defined by
\begin{equation}\label{ch-22}
F^*_\alpha(x,u,v,u_{(1)},m_{(1)},...,u_{(s)},m_{(s)})=E_{u^\alpha}(m^\beta F_\beta)=0,~\alpha=1,2,...,N,
\end{equation}
where
$E_{u^\alpha}=\frac{\partial }{{\partial u^\alpha  }} + \sum\limits_{\mu = 1}^s {( - )^\mu D_{i_1 } ,...,D_{i_\mu } \frac{\partial }{{\partial u_{i_1 ,...,i_\mu }^\alpha  }}}$ denotes the Euler-Lagrange operator, $m=m(x)$ is a new dependent variable $m=(m^1,m^2,...,m^N)$.

\textbf{Theorem 1 }:\cite{Ibragimov2}. The system consisting of Eqs.(\ref{ch-20}) and their adjoint Eqs.(\ref{ch-22})
\begin{equation}\label{ch-23}
\left\{ {\begin{array}{*{20}c}
   {F_\alpha  (x,u,u_{(1)} ,...,u_{(s)} ) = 0,{\rm{~~~~~~~~~~~~~~}}}  \\
   {F_\alpha ^* (x,u,m,u_{(1)} ,m_{(1)} ,...,u_{(s)} ,m_{(s)} ) = 0,}  \\
\end{array}} \right.
\end{equation}
has a formal Lagrangian
\begin{equation}\label{ch-24}
L=m^\beta F_\beta  (x,u,u_{(1)} ,...,u_{(s)} ).
\end{equation}

\textbf{Theorem 2}:\cite{Ibragimov2} Any Lie point, Lie-B\"{a}cklund and non-local symmetry
\begin{equation}\label{ch-25}
V=\xi^i \frac{\partial }{\partial x^i}+\eta^\alpha \frac{\partial }{\partial u^\alpha},
\end{equation}
of Eqs.(\ref{ch-20})provides a conservation law $D_i(T^i)=0$ for the system comprising Eqs.(\ref{ch-20}) and its adjoint Eqs.(\ref{ch-22}). The conserved vector is given by
\begin{equation}\label{ch-26}
T^i  = \xi ^i L + W^\alpha  E_{u_i^\alpha  } (L) + D_j (W^\alpha  )E_{u_{ij}^\alpha  } (L) + D_j D_k (W^\alpha  )E_{u_{ijk}^\alpha  } (L) + ...,
\end{equation}
where $W^\alpha$ is the Lie characteristic function
\begin{center}
    $W^\alpha=\eta^\alpha-\xi^j u^\alpha_j,$
\end{center}
and $L$ is determined by Eqs.(\ref{ch-24}).

Next, we construct the conservation laws Eqs(\ref{ch-2}). Let us consider the prolonged system, one can obtain the following results from the above,
\begin{center}
$\begin{array}{l}
 \xi ^1  = c_3 ,\xi ^2  = c_4 ,\eta ^1  = c_2 \varphi ^2  + c_1 u, \eta ^2  = c_2 \phi ^2  - c_1 v,\\
\eta ^3  = \frac{{(2c_2 f + c_1  + c_5 )\varphi }}{2}, \eta ^4  = \frac{{( - 2c_2 f + c_1  - c_5 )\phi }}{2},\eta ^5  = c_2 f^2  + c_5 f + c_6 .\\
 \end{array}$
\end{center}
and
\begin{center}
$\begin{array}{l}
 L=m^1 (iu_t  + \alpha (u_{xx}  - 2u^2 v) + i\beta (u_{xxx}  - 6uvu_x )) + m^2 (iv_t  - \alpha (v_{xx}  - 2v^2 u) + i\beta (v_{xxx}  - 6uvv_x ))+ m^3 ( - i\lambda \phi  + u\psi ) \\
  + m^4 (v\phi  + i\lambda \psi ) + m^5 (a\phi  + b\psi ) + m^6 (c\phi  - a\psi ) + m^7 (\varphi \phi )+ m^8 ( - \beta \varphi ^2 v_x  + 12\beta \lambda ^2 \varphi \phi + 4\lambda \alpha \varphi \phi  - \beta \phi ^2 u_x + 2\beta \varphi \phi uv\\
+4i\lambda \beta \phi ^2 u - 4i\lambda \beta \varphi ^2 v + i\alpha \phi ^2 u - i\alpha \varphi ^2 v) \\
 \end{array}$
\end{center}
and $a,b,c$ are determined by Eqs.(\ref{ch-3-1}).

Using the theorem 2, one can get the following results by calculation,
\begin{equation}\label{ch-25}
\begin{array}{l}
 T^1 = m^1 (c_3 \beta u_{xxx}  + 2i\alpha c_3 u^2 v + c_2 \phi ^2  - c_4 u_x  + 2c_1 u - c_3 i\alpha u_{xx}  - 6c_3 \beta uvu_x ) + m^2 ( - 6c_3 \beta uvv_x  + c_2 \psi ^2  - c_4 v_x  \\
  - 2c_1 v + c_3 \beta v_{xxx}  - 2ic_3 \alpha uv^2  + c_3 \alpha iv_{xx} ) + m^4 (c_3 i\alpha \phi uv + c_3 \beta u\phi v_x  - 2c_3 \beta u^2 v\psi  - 4c_3 \beta \lambda ^2 \psi u - c_4 \psi u - 2c_3 \alpha \lambda u\psi \\
 + c_3 \beta \psi u_{xx}  + c_4 i\lambda \phi  - c_3 i\alpha \psi u_x  - c_3 \beta \phi vu_x  + 2ic_3 \beta \lambda uv\phi  + 2ic_3 \alpha \phi \lambda ^2 c_2 f\phi  + 4ic_3 \beta \phi \lambda ^3  - 2ic_3 \beta \lambda \psi u_x  + c_5 \phi  + c_1 \phi )\\
 + m^6 ( - c_4 i\lambda \psi  - 2ic_3 \beta \lambda \psi uv + c_3 \beta v\psi u_x  - 4c_3 \beta \phi v\lambda ^2  + c_5 \psi  + c_2 f\psi  + 2ic_3 \beta \lambda \phi v_x  - c_3 i\alpha uv\psi  - c_4 v\phi  + c_3 \beta \phi v_{xx}\\
  - 2c_3 \alpha \lambda \phi v- 2c_3 \beta \phi uv^2  - 4ic_3 \beta \lambda ^3 \psi  + c_3 i\alpha \phi v_x  - c_3 \beta u\psi v_x  - 2c_3 i\alpha \lambda ^2 \psi  - c_1 \psi ) + m^8 ( - 2c_3 \beta \phi uv\psi - 4ic_3 \beta \lambda u\psi ^2 \\
 - c_4 \phi \psi  - 12c_3 \beta \lambda ^2 \phi \psi  - 4c_3 \alpha \lambda \phi \psi  - c_3 i\alpha u\psi ^2  + c_2 f^2  + 2c_5 f + c_6  + c_3 \beta \phi ^2 v_x  + c_3 \beta \psi ^2 u_x  + 4ic_3 \beta \lambda v\phi ^2  + c_3 i\alpha v\phi ^2 )\\
 \end{array}
\end{equation}
and
\begin{equation}\label{ch-26}
\begin{array}{l}
 T^2 = m^7 (c_2 f^2  + 2c_5 f + c_6  - c_3 f_t  - c_4 \psi \phi ) + m^5 ( - c_3 \psi _t  + c_5 \psi  - c_1 \psi  - c_4 v\phi  - c_4 i\lambda \psi  + c_2 f\psi ) + m^8 (2c_2 \beta \psi ^2 \phi ^2  \\
 c_3 \beta \phi ^2 v_t  + 2c_1 \beta u\psi ^2  - 2c_1 \beta v\phi ^2  + c_4 f_t  - 2c_4 \beta \psi \phi uv - 4ic_4 \beta \lambda u\psi ^2  + 4ic_4 \beta \lambda v\phi ^2  + c_4 i\alpha v\phi ^2  - 12c_4 \beta \lambda ^2 \psi \phi  \\
  - 4c_4 \lambda \alpha \psi \phi  - c_4 i\alpha u\psi ^2  - c_3 \beta \psi ^2 u_t ) + m^6 (c_4 \psi _t  - 4ic_1 \beta \lambda v\phi  - c_4 i\alpha uv\psi  - 2ic_3 \beta \lambda \phi v_t  + 2ic_2 \beta \lambda \psi ^2 \phi  + c_2 \beta v\psi \phi ^2  \\
  + 4c_1 \beta uv\psi  + c_2 i\alpha \psi ^2 \phi  - c_3 \beta v\psi u_t  + c_3 \beta u\psi v_t  + 2c_2 \beta \psi \phi \psi _x  - 2c_4 \beta uv^2 \phi  - 4c_4 \beta \lambda ^2 v\phi  - 2c_4 \alpha \lambda v\phi  - 2ic_1 \alpha v\phi  \\
  - i\alpha c_3 \phi v_t  - 4ic_4 \beta \lambda ^3 \psi  - 2ic_4 \lambda ^2 \alpha \psi  - 2ic_4 \lambda \beta uv\psi  - c_2 \beta \psi ^2 \phi _x  + c_3 \beta v_t \phi _x  + c_4 \beta v_x \phi _x  - 2c_1 \beta v_x \phi  - c_2 \beta u\psi ^3  \\
  - c_3 \beta \phi v_{xt} ) + m^4 (c_4 \phi _t  - 4c_1 \beta uv\phi  + c_2 \beta u\psi ^2 \phi  - c_3 \beta u\phi v_t  + 2c_2 \beta \phi \psi \phi _x  - 2c_4 \beta u^2 v\psi  - 4c_4 \beta \lambda ^2 u\psi  - 2c_4 \lambda \alpha u\psi  \\
  + c_3 \beta v\phi u_t  + c_3 i\alpha \psi u_t  - c_2 i\alpha \psi \phi ^2  - 2c_1 i\alpha u\psi  + 4ic_4 \beta \lambda ^3 \phi  + 2ic_4 \alpha \lambda ^2 \phi  + ic_4 \alpha uv\phi  + 2ic_3 \beta \lambda \psi u_t  - 2ic_2 \beta \lambda \psi \phi ^2  \\
  - 4ic_1 \beta \lambda \psi u + 2ic_4 \beta \lambda uv\phi  - c_3 \beta \psi u_{xt}  - c_2 \beta \phi ^2 \psi _x  - 2c_1 \beta u\psi _x  + c_3 \beta \psi _x u_t  + c_4 \beta \psi _x u_x  + 2c_1 \beta \psi u_x  - c_2 \beta \phi ^3 v) \\
  + m^2 (2c_2 \beta \psi \psi _{xx}  - 6c_2 \beta uv\psi ^2  - 2c_1 \beta v_{xx}  + 12c_1 \beta uv^2  - ic_3 \alpha v_{xt}  + 2ic_2 \alpha \psi \psi _x  - 2ic_4 \alpha uv^2  + c_4 v_t  + 6c_3 \beta uvv_t  \\
  - 2ic_1 \alpha v_x  + 2c_2 \beta \psi _x^2  - c_3 \beta v_{xxt} ) + m^1 ( - 2ic_1 \alpha u_x  + c_4 u_t  + ic_3 \alpha u_{xt}  + 2c_1 \beta u_{xx}  + 2c_2 \beta \phi \phi _{xx}  + 6c_3 \beta uvu_t  - 12c_1 \beta vu^2  \\
  - 2ic_2 \alpha \phi \phi _x  + 2c_2 \beta \phi _x^2  - c_3 \beta u_{xxt}  - 6c_2 \beta uv\phi ^2  + 2ic_4 \alpha u^2 v) + m^3 ( - c_4 u\psi  + c_5 \phi  + ic_4 \lambda \phi  + c_1 \phi  + c_2 \phi f - c_3 \phi _t ) -  \\
 2c_1 \beta u_x m^1_x  + 2c_1 \beta v_x m^2_x  + ic_2 \alpha \phi ^2 m^1_x  - c_2 \beta \psi \phi ^2 m^4_x  - 2c_1 \beta u\psi m^4_x  - ic_4 \alpha u_x m^1_x  - ic_2 \alpha \psi ^2 m^2_x + 2i\alpha c_1 v m^2_x  +  \\
 2i\alpha c_1 um^1_x  + c_4 \beta \psi u_x m^4_x  + ic_4 \alpha v_x m^2_x  + ic_3 \alpha v_t m^2_x  - c_2 \beta \psi ^2 \phi v_x m^6_x  + c_4 \beta \phi v_x m^6_x  + c_3 \beta \phi v_t m^6_x  + 2c_1 \beta \phi vm^6_x  \\
  - 2c_2 \beta \phi \phi _x m^1_x  - 2c_2 \beta \psi \psi _x m^2_x  - c_3 \beta v_t m^2_{xx}  + c_3 \beta v_{xt} m^2_x  - c_3 \beta u_t m^1_{xx}  + c_3 \beta u_{xt} m^1_{x}  + c_2 \beta \phi ^2 m^1_{xx}  + 2c_1 \beta um^1_{xx}  \\
  - c_4 \beta u_x m^1_{xx}  + c_2 \beta \psi ^2 m^2_{xx}  - c_4 \beta v_x m^2_{xx}  - 2c_1 \beta vm^2_{xx}  + c_4 \beta m^1_{x} u_{xx}  + c_4 \beta m^2_{x} v_{xx}  - ic_3 \alpha m^1_{x} u_t  + c_3 \beta \psi m^4_{x} u_t  \\
 \end{array}
\end{equation}
where $m^1_t,m^2_t,m^3_x,m^5_x,m^7_x$ can be obtained by using (\ref{ch-22}) and (\ref{ch-24}). Because the formulas are complicated, so here we omit.

Through the verification, $T^1,T^2$ satisfy the equation(\ref{ch-21}),i.e.
\begin{equation}\label{ch-27}
D_t(T^1)+D_x(T^2)=0,
\end{equation}
thus, Eqs.(\ref{ch-25}),(\ref{ch-26}) define the corresponding components of a non-local conservation law for the system of Eqs. (\ref{ch-1}).

\textbf{Remark1}: For the coupled Hirota equations (\ref{ch-1}), we get its conservation laws by making use of the explicit solution of the adjoint equations of Eqs. (\ref{ch-1}). But the number of the adjoint equations is less than the number of variables, so $T^1,T^2$ have arbitrary functions, we can conclude that Eqs.(\ref{ch-1}) have infinitely many conservation laws.

\section{Summary and Discussion}\label{conclusion}
In this paper, the nonlocal symmetry of coupled Hirota equations is obtained by using the lax pair and localized by introducing an auxiliary dependent variable. Then, the primary nonlocal symmetry is equivalent to a Lie point symmetry of a prolonged system. On the basis of this system, the one-dimensional subalgebras of a Lie algebra have been classified and the reductions of coupled Hirota equations are given out by using the associated vector fields. On the one hand, non-trivial nonlocal conservation laws of the coupled Hirota equation are obtained by means of these formulas. For general evolution equation, the nonlocal conservation laws may be not enough to guarantee the integrability. However, it must be useful to help to understand the special solutions of the nonlinear systems.

This method would be possible to extend to many other interesting integrable models. However, there is not a universal way to estimate what kind of nonlocal symmetries can be localized to some related prolonged system, so there are still many questions worth to study. Moreover, one can construct infinitely many nonlocal symmetries by introducing some internal parameters from the seed symmetry and infinitely many nonlocal conservation laws of the completely integrable finite-dimensional systems. Above topics will be discussed in the future series research works.

\section*{Acknowledgment}
%We would like to express our sincere thanks to Professor *** and other members of our discussion group for their valuable comments.
%This work is supported by .

\begin{flushleft}
\textbf{References}
\end{flushleft}

%% Authors are advised to submit their bibtex database files. They are
%% requested to list a bibtex style file in the manuscript if they do
%% not want to use model1-num-names.bst.

%% References without bibTeX database:

% \begin{thebibliography}{00}

%% \bibitem must have the following form:
%%   \bibitem{key}...
%%

% \bibitem{}

% \end{thebibliography}


\begin{thebibliography}{99}
%\itemsep=-2pt
%
\bibitem{Lie1} S. Lie, \"{U}ber die Integration durch bestimmte Integrale von einer Klasse linearer partieller Differential gleichungen, Arch. Math. 328 (1881).
\bibitem{Ovsiannikov1} L.V. Ovsiannikov, \emph{Group Analysis of Differential Equations} (New York: Academic, 1982).
\bibitem{Ibragimov1} N.H. Ibragimov, \emph{Transformation Groups Applied to Mathematical Physics} (Boston, MA: Reidel, 1985).
\bibitem{Bluman1} G.W. Bluman and S. Kumei, \emph{Symmetries and Differential Equations} (Springer-Verlag, New York 1989).
\bibitem{Chen1} X.R. Hu, F. Huang and Y. Chen, Symmetry reductions and exact solutions of the (2+1)-dimensional navier-stokes equations, Z. Naturforsch. 65a(2010) 1-7.
\bibitem{Liu1}H.Z. Liu, J.B. Li, Symmetry reductions, dynamical behavior and exact explicit solutions to the Gordon types of equations, J. Comput. Appl. Math., 257 (2014), 144-156.
\bibitem{Olver1} P.J. Olver, \emph{Applications of Lie Groups to Differential Equations} (Berlin: Springer, 1986).
\bibitem{Akhatov} I.S. Akhatov, R.K. Gazizov and N.K. Ibragimov, Nonlocal symmetries. Heuristic approach, J. Sov. Math. 55(1991) 1401-1450.
\bibitem{Bluman} G.W. Bluman, A.F. Cheviakov and S.C. Anco, \emph{Applications of Symmetry Methods to Partial Differential Equations} (Springer New York, 2010).
\bibitem{Lou} S.Y. Lou and X.B. Hu, Non-local symmetries via Darboux transformations, J. Phys. A: Math. Gen. 30(1997) L95-L100.

\bibitem{Galas1} F. Galas, New  non-local symmetries with pseudopotentials, J. Phys. A: Math. Gen. 25(1992) L981-L986.
\bibitem{Lou1} S.Y. Lou, X.R. Hu and Y. Chen, Nonlocal symmetries related to B\"{a}cklund transformation and their applications, J. Phys. A: Math. Theor. 45(2012) 155209-155209.
\bibitem{Hu1} X.R. Hu, S.Y. Lou and Y. Chen, Explicit solutions from eigenfunction symmetry of the Korteweg-deVries equation, Phys. Rev. E 85(2012) 056607 1-8.

\bibitem{David1} D. David, N. Kamran, D. Levi and P. Winternitz, Symmetry reduction for the Kadomtsev¨CPetviashvili equation using a loop algebra. J. Math. Phys. 27 (1986), 1225-1237.
\bibitem{Qu1} C.Z. Qu, Group classification of a class of coupled equations, Int. J. Thero. Phys., 34(1995) 2015-2030.
\bibitem{Liu2} H.Z. Liu, J.B. Li, L. Liu, Y. Wei, Group classifications, optimal systems and exact solutions to the generalized Thomas equations, J. Math. Anal. Appl., 383(2011) 400-408.
\bibitem{Hu2} X.R. Hu, Y.Q. Li and Y. Chen, A direct algorithm of one-dimensional optimal system for the group invariant solutions, arXiv:1411.3797v1.
\bibitem{Noether1} E. Noether, Invariante variationsprobleme, Nachr. K\"{o}nig. Gesell. Wissen., G\"{o}ttingen, Math.-Phys. Kl. Heft 2(1918) 235-257.
\bibitem{Steudel1} H. Steudel, Uber die zuordnung zwischen invarianzeigenschaften und erhaltungssatzen, Z. Naturforsch. 17a(1962) 129-132.
\bibitem{Anco1} S.C. Anco, G.W. Bluman, Direct construction method for conservation laws of partial differential equations. Part I: Examples of conservation law classifications, Eur. J. Appl. Math. 13(2002) 545-566.
\bibitem{Lou2} S.Y. Lou, Nonlocal conservation laws and related B\"{a}cklund transformations via reciprocal transformations,  arXiv:1402.7231v2.
\bibitem{Ibragimov2} N.H. Ibragimov, A new conservation theorem, J. Math. Anal. Appl., 333(2007) 311-328.


\bibitem{Ablowitz1} M.J. Ablowitz, D.J. Kaup, A.C. Newell, and H.Segur, Nonlinear-evolution equations of physical significance, Phys. Rev. Lett. 31(1973) 125-127.
\bibitem{Ankiewicz1} A. Ankiewicz, N. Akhmediev, and J. M. Soto-Crespo, Discrete rogue waves of the Ablowitz-Ladik and Hirota equations, Phys. Rev. E, 82(2010) 026602.

\bibitem{Zhang1} Y. Zhang, K.H. Dong, and R.J. Jin, The Darboux transformation for the coupled Hirota equation, AIP Conf. Proc.1562, 249(2013) 249-256.

\bibitem{Xin1} X.P. Xin, Y. Chen, A Metho d to Construct the Nonlo cal Symmetries of Nonlinear Evolution Equations, Chin. Phys. Lett. 30(2013) 100202.



%\bibitem{Bluman} G.W. Bluman, Temuerchaolu, and R. Sahadevan, Local and nonlocal symmetries for nonlinear telegraph equation, J. Math. Phys. 46(2005) 023505 1-12.
%\bibitem{Bluman} G.W. Bluman, G.J. Reid, S.Kumei, New classes of symmetries for partial differential equations, J. Math. Phys. 29 (1988) 806.




%%%%%%%%%%%%%%%%%%%%%%%%%%%%%%%%%%%%








\end{thebibliography}
\end{document}